\documentclass[prd,a4paper,twocolumn]{revtex4}%
\usepackage{amsfonts}
\usepackage{amsmath}
\usepackage{amssymb}
\usepackage{graphicx}%
\newtheorem{lemma}{Lemma}

\newtheorem{remark}{Remark}

\begin{document}
\title{Linking electroweak and gravitational generators}
\author{John Fredsted}
\email{physics@johnfredsted.dk}
\affiliation{Soeskraenten 22, Stavtrup, DK-8260 Viby J., Denmark}

\begin{abstract}
Using complexified quaternions, an intriguing link between generators of two
different and surprisingly \textit{commuting} four-dimensional representations
of the $\mathrm{SU}\left(  2\right)  \times\mathrm{U}\left(  1\right)  $ Lie
group, and generators of two four-dimensional spin $\frac{1}{2}$
representations of the $\mathrm{Spin}\left(  3,1\right)  $ Lie group is
established: the former generators completely determine the latter ones, and
cross-combined they constitute two different, but closely related,
four-dimensional representations of $\mathrm{Spin}\left(  3,1\right)
\times\mathrm{SU}\left(  2\right)  \times\mathrm{U}\left(  1\right)  $. These
representations are used to construct a $\mathrm{Spin}\left(  3,1\right)
\times\mathrm{SU}\left(  2\right)  \times\mathrm{U}\left(  1\right)  $ gauge
invariant Lagrangian, containing two four-spinors consisting not as usual of
Weyl two-spinors of opposite helicity and equal weak isospin, but instead of
Weyl two-spinors of opposite weak isospin and equal helicity, a construction
which arises naturally from the mathematical formalism itself. A possible
future generalization, using complexified octonions, is discussed.

\end{abstract}
\maketitle

\section{Introduction and main result}

The quest for unification of the fundamental forces of Nature began with the
unification of electricity and magnetism, by Maxwell, resulting in the
electromagnetic force \cite{Jackson}. A century later this force was united
with the weak nuclear force, resulting in the electroweak force
\cite{Weinberg}. Even though there have been promising theoretical
propositions of unification of this force with the strong nuclear force, none
of these have been experimentally verified. Gravity \cite{MTW} is in its very
own category, stubbornly refusing to join the quantum-party of unification,
the currently most promosing, though highly speculative, theoretical
proposition being string/M-theory \cite{GSW,Kiritsis}.

This article makes no claim of any unification. More modestly, it is the
purpose of this article to point out, and utilize, an intriguing link between
generators of two different and surprisingly \textit{commuting}, see Eq.
(\ref{Eq:GamsCommute}), four-dimensional representations of the $\mathrm{SU}%
\left(  2\right)  \times\mathrm{U}\left(  1\right)  $ Lie group (relevant for
the electroweak force), and generators of two four-dimensional spin $\frac
{1}{2}$ representations of the $\mathrm{Spin}\left(  3,1\right)  $ Lie group
(relevant for the interaction, described in terms of a minimal spin
connection, of the gravitational field and spinor fields): the former
generators completely determine the latter ones, see Eq. (\ref{Eq:GensEtaCom}%
), and cross-combined they constitute two different, but closely related (by
complex conjugation), four-dimensional representations of $\mathrm{Spin}%
\left(  3,1\right)  \times\mathrm{SU}\left(  2\right)  \times\mathrm{U}\left(
1\right)  $, see Sec. \ref{Sec:GensAll}. Mathematically, this link between
generators is (most directly) established using complexified quaternions, also
called biquaternions. Physically, this link is established by grouping
together in two four-spinors not the usual Weyl two-spinors of opposite
helicity and equal weak isospin, but instead Weyl two-spinors of opposite weak
isospin and equal helicity, a construction which arises naturally from the
mathematical formalism itself. The main result of the article is the
$\mathrm{Spin}\left(  3,1\right)  \times\mathrm{SU}\left(  2\right)
\times\mathrm{U}\left(  1\right)  $ gauge invariant Lagrangian, Eq.
(\ref{Eq:Lagrangian}).

No formal introduction to the quaternions will be given, the reader kindly
being referred to for instance Refs. \cite{Lounesto,Lambek}. Neither will any
formal introduction to the (complexified) octonions, or the even more general
composition algebras, be given, the reader kindly being referred to the
literature: For short reviews of the octonions, see Refs. \cite{Gunaydin and
Gursey,Dundarer and Gursey,Dundarer Gursey and Tze,Bakas et al.}. For a
comprehensive review of the octonions, see Ref. \cite{Baez}. For a monograph
on octonions and other nonassociative algebras, see Ref. \cite{Okubo}. In
particularly, for a monograph on composition algebras, a class to which both
the complex quaternions and complex octonions belong (note that they are not
division algebras, even though the quaternions and the octonions themselves
are), see Ref. \cite{Springer and Veldkamp}.

The paper is organized as follows: Sec. \ref{Section:Notation and conventions}
introduces the necessary notation and conventions used. Sec.
\ref{Section:Setup} sets up the main machinery needed. Sec.
\ref{Section:Lagrangian} contains the main result of the paper; the
$\mathrm{Spin}\left(  3,1\right)  \times\mathrm{SU}\left(  2\right)
\times\mathrm{U}\left(  1\right)  $ gauge invariant Lagrangian, Eq.
(\ref{Eq:Lagrangian}). Sec. \ref{Section:Discussion} discusses various notable
features of this Lagrangian, and points to a future generalization, using
complexified octonions. There are two appendices: Appendix
\ref{Appendix:Identities} contains various useful identities valid for any
composition algebra, a class to which both the complexified quaternions and
complexified octonions belong. Appendix \ref{Appendix:Proofs} contains the
proofs of most of the assertions of Sec. \ref{Section:Setup}.

\section{Notation and conventions\label{Section:Notation and conventions}}

The set of complexified quaternions is denoted $\mathbb{C}\otimes\mathbb{H}$,
equal to $\mathbb{H}\otimes\mathbb{C}$ because the complex numbers
$\mathbb{C}$ and the quaternions $\mathbb{H}$ are assumed to commute. The
imaginary unit of $\mathbb{C}$ is denoted $\mathrm{i}$, obeying $\mathrm{i}%
^{2}=-1$, of course. The imaginary units of $\mathbb{H}$ are denoted
$\mathrm{e}_{i}=\left(  \mathrm{e}_{1},\mathrm{e}_{2},\mathrm{e}_{3}\right)
$, obeying $\mathrm{e}_{i}\mathrm{e}_{j}=-\delta_{ij}+\varepsilon_{ij}{}%
^{l}\mathrm{e}_{l}$, where $\varepsilon_{ijk}$ is the Levi-Civita symbol with
$\varepsilon_{123}=+1$. The basis for $\mathbb{C}\otimes\mathbb{H}$ (over
$\mathbb{C}$) is taken as $\mathrm{e}_{a}=\left(  \mathrm{e}_{0}%
,\mathrm{e}_{i}\right)  =\left(  \mathrm{i},\mathrm{e}_{i}\right)  $.

Latin indices from the beginning of the alphabet run from $0$ to $3$, and are
raised and lowered with $\eta^{ab}$ and $\eta_{ab}$, respectively, $\eta_{ab}$
being the Minkowski metric. Latin indices from the middle of the alphabet,
beginning at $i$, run from $1$ to $3$, and are raised and lowered with
$\delta^{ij}$ and $\delta_{ij}$, respectively. Greek indices run from $0$ to
$3$, and are raised and lowered with $g^{\mu\nu}$ and $g_{\mu\nu}$,
respectively, $g_{\mu\nu}$ being the metric of curved spacetime. The Einstein
summation convention is adhered to throughout.

Let $c^{a}\in\mathbb{C}$. Complex conjugation is the involution $\cdot^{\ast
}:\mathbb{C}\otimes\mathbb{H\rightarrow C}^{\ast}\otimes\mathbb{H}$ defined by
$\left(  c^{a}\mathrm{e}_{a}\right)  ^{\ast}=-\left(  c^{0}\right)  ^{\ast
}\mathrm{e}_{0}+\left(  c^{i}\right)  ^{\ast}\mathrm{e}_{i}$, and quaternionic
conjugation is the involution $\overline{\cdot}:\mathbb{C}\otimes
\mathbb{H\rightarrow C}\otimes\overline{\mathbb{H}}$ defined by $\overline
{c^{a}\mathrm{e}_{a}}=c^{0}\mathrm{e}_{0}-c^{i}\mathrm{e}_{i}$. Note that
$\overline{\mathrm{e}}_{a}^{\ast}=-\mathrm{e}_{a}$.

The bilinear inner product $\left\langle \cdot,\cdot\right\rangle :\left(
\mathbb{C}\otimes\mathbb{H}\right)  ^{2}\mathbb{\rightarrow C}$ is defined by%
\begin{equation}
2\left\langle x,y\right\rangle =x\overline{y}+y\overline{x}\equiv\overline
{x}y+\overline{y}x. \label{Eq:ipDef}%
\end{equation}
Note that $\left\langle \mathrm{e}_{a},\mathrm{e}_{b}\right\rangle =\eta_{ab}$.

The set of $n$-dimensional square matrices over some field $\mathbb{F}$ is
denoted $\mathrm{M}\left(  n,\mathbb{F}\right)  $. The $n$-dimensional
identity matrix is denoted $\mathbf{1}_{n}$, and the $n$-dimensional matrix
with zero entries only is denoted $\mathbf{0}_{n}$. The socalled
'eta-transpose' $\cdot^{\eta}:\mathrm{M}\left(  4,\mathbb{C}\right)
\rightarrow\mathrm{M}\left(  4,\mathbb{C}\right)  $ is defined by%
\[
\mathbf{A}^{\eta}=\mathbf{\eta A}^{\mathrm{T}}\mathbf{\eta,}%
\]
where $\mathbf{\eta}$ is the Minkowski metric as a matrix; $\left(
\mathbf{\eta}\right)  ^{a}{}_{b}=\eta_{ab}=\eta^{ab}$. In terms of this
'eta-transpose', the two (anti)commutator-like [even though they do not have
all the properties of the usual (anti)commutator] brackets $\left[
\cdot,\cdot\right]  _{\eta\pm}:\mathrm{M}\left(  4,\mathbb{C}\right)
\rightarrow\mathrm{M}\left(  4,\mathbb{C}\right)  $, are defined by%
\[
\left[  \mathbf{A},\mathbf{B}\right]  _{\eta\pm}=\mathbf{A}^{\eta}%
\mathbf{B}\pm\mathbf{B}^{\eta}\mathbf{A}.
\]

\section{Setup\label{Section:Setup}}

For analytical proofs of various assertions of this section, see Appendix
\ref{Appendix:Proofs}.

\subsection{Generators of $\mathrm{SU}\left(  2\right)  \times\mathrm{U}%
\left(  1\right)  $}

Define the matrices $\mathbf{\Gamma}_{L|a},\mathbf{\Gamma}_{R|a}\in
\mathrm{M}\left(  4,\mathbb{C}\right)  $ by
\begin{subequations}
\label{Eq:GamsDef}%
\begin{align}
\left(  \mathbf{\Gamma}_{L|a}\right)  _{cd}  &  =\left\langle \mathrm{e}%
_{c},\mathrm{e}_{a}\mathrm{e}_{d}\right\rangle ,\label{Eq:GamsDefL}\\
\left(  \mathbf{\Gamma}_{R|a}\right)  _{cd}  &  =\left\langle \mathrm{e}%
_{c},\mathrm{e}_{d}\mathrm{e}_{a}\right\rangle , \label{Eq:GamsDefR}%
\end{align}
where $L$ and $R$ refer to whether $\mathrm{e}_{a}$ is multiplied from the
left or the right. They obey (note the mixed positions of $L$ and $R$)
\end{subequations}
\begin{subequations}
\label{Eq:GamsInvolution}%
\begin{align}
\mathbf{\Gamma}_{L|a}^{\ast}  &  =-\mathbf{\Gamma}_{R|a},\quad\mathbf{\Gamma
}_{R|a}^{\ast}=-\mathbf{\Gamma}_{L|a},\label{Eq:GamsComplexConj}\\
\mathbf{\Gamma}_{L|a}^{\mathrm{T}}  &  =+\mathbf{\Gamma}_{R|a},\quad
\mathbf{\Gamma}_{R|a}^{\mathrm{T}}=+\mathbf{\Gamma}_{L|a}%
,\label{Eq:GamsTransposition}\\
\mathbf{\Gamma}_{L|a}^{\dagger}  &  =-\mathbf{\Gamma}_{L|a},\quad
\mathbf{\Gamma}_{R|a}^{\dagger}=-\mathbf{\Gamma}_{R|a}.
\label{Eq:GamsHermitianConj}%
\end{align}
Also, they obey the Lie algebra (where $X$ denotes either $L$ or $R$, a
shorthand notation frequently used below)
\end{subequations}
\begin{subequations}
\label{Eq:GamsLieAlgebra}%
\begin{align}
-2\varepsilon_{ij}{}^{k}\mathbf{\Gamma}_{L|k}  &  =\left[  \mathbf{\Gamma
}_{L|i},\mathbf{\Gamma}_{L|j}\right]  ,\label{Eq:GamsLieAlgebraL}\\
+2\varepsilon_{ij}{}^{k}\mathbf{\Gamma}_{R|k}  &  =\left[  \mathbf{\Gamma
}_{R|i},\mathbf{\Gamma}_{R|j}\right]  ,\label{Eq:GamsLieAlgebraR}\\
\mathbf{0}_{4}  &  =\left[  \mathbf{\Gamma}_{X|0},\mathbf{\Gamma}%
_{X|i}\right]  , \label{Eq:GamsLieAlgebraX}%
\end{align}
so that $\mathrm{i}\mathbf{\Gamma}_{L|a}$ and $\mathrm{i}\mathbf{\Gamma}%
_{R|a}$ (note the $\mathrm{i}$'s) each constitute hermitian generators of two
different, but closely related (by complex conjugation), four-dimensional
representations of $\mathrm{SU}\left(  2\right)  \times\mathrm{U}\left(
1\right)  $. In fact, due to the surprising relation (which originally
prompted this research)%
\end{subequations}
\begin{equation}
\mathbf{0}_{4}=\left[  \mathbf{\Gamma}_{L|a},\mathbf{\Gamma}_{R|b}\right]  ,
\label{Eq:GamsCommute}%
\end{equation}
they constitute two \textit{commuting} representations.

Furthermore, they obey the anticommutator-like relation%
\begin{equation}
2\eta_{ab}\mathbf{1}_{4}=\left[  \mathbf{\Gamma}_{X|a},\mathbf{\Gamma}%
_{X|b}\right]  _{\eta+}\equiv\mathbf{\Gamma}_{X|a}^{\eta}\mathbf{\Gamma}%
_{X|b}+\mathbf{\Gamma}_{X|b}^{\eta}\mathbf{\Gamma}_{X|a},
\label{Eq:GamsEtaAntiCom}%
\end{equation}
which is the reason for the choice of '$\mathbf{\Gamma}$' as the designating
letter, $\mathbf{\Gamma}_{L|a}$ and $\mathbf{\Gamma}_{R|a}$ being reminiscent
of the usual Clifford gamma matrices.

\subsection{Generators of $\mathrm{Spin}\left(  3,1\right)  $}

Define the matrices $\mathbf{\Sigma}_{L|ab},\mathbf{\Sigma}_{R|ab}%
\in\mathrm{M}\left(  4,\mathbb{C}\right)  $ by
\begin{subequations}
\label{Eq:GensDef}%
\begin{align}
4\mathrm{i}\left(  \mathbf{\Sigma}_{L|ab}\right)  _{cd}  &  =\left\langle
\mathrm{e}_{a}\mathrm{e}_{c},\mathrm{e}_{b}\mathrm{e}_{d}\right\rangle
-\left\langle \mathrm{e}_{a}\mathrm{e}_{d},\mathrm{e}_{b}\mathrm{e}%
_{c}\right\rangle ,\label{Eq:GensDefL}\\
4\mathrm{i}\left(  \mathbf{\Sigma}_{R|ab}\right)  _{cd}  &  =\left\langle
\mathrm{e}_{c}\mathrm{e}_{a},\mathrm{e}_{d}\mathrm{e}_{b}\right\rangle
-\left\langle \mathrm{e}_{d}\mathrm{e}_{a},\mathrm{e}_{c}\mathrm{e}%
_{b}\right\rangle , \label{Eq:GensDefR}%
\end{align}
where $L$ and $R$ refer to whether $\mathrm{e}_{a}$ and $\mathrm{e}_{b}$ are
multiplied from the left or the right. They are related to $\mathbf{\Gamma
}_{L|a}$ and $\mathbf{\Gamma}_{L|b}$ by the commutator-like relations%
\end{subequations}
\begin{equation}
4\mathrm{i}\mathbf{\Sigma}_{X|ab}=\left[  \mathbf{\Gamma}_{X|a},\mathbf{\Gamma
}_{X|b}\right]  _{\eta-}\equiv\mathbf{\Gamma}_{X|a}^{\eta}\mathbf{\Gamma
}_{X|b}-\mathbf{\Gamma}_{X|b}^{\eta}\mathbf{\Gamma}_{X|a}.
\label{Eq:GensEtaCom}%
\end{equation}
They obey (note the mixed positions of $L$ and $R$)
\begin{subequations}
\label{Eq:GensInvolution}%
\begin{align}
\mathbf{\Sigma}_{L|ab}^{\ast}  &  =-\mathbf{\Sigma}_{R|ab},\quad
\mathbf{\Sigma}_{R|ab}^{\ast}=-\mathbf{\Sigma}_{L|ab}%
,\label{Eq:GensComplexConj}\\
\mathbf{\Sigma}_{L|ab}^{\mathrm{T}}  &  =-\mathbf{\eta\Sigma}_{L|ab}%
\mathbf{\eta},\quad\mathbf{\Sigma}_{R|ab}^{\mathrm{T}}=-\mathbf{\eta\Sigma
}_{R|ab}\mathbf{\eta},\label{Eq:GensTransposition}\\
\mathbf{\Sigma}_{L|ab}^{\dagger}  &  =+\mathbf{\eta\Sigma}_{R|ab}\mathbf{\eta
},\quad\mathbf{\Sigma}_{R|ab}^{\dagger}=+\mathbf{\eta\Sigma}_{L|ab}%
\mathbf{\eta}. \label{Eq:GensHermitianConj}%
\end{align}
Also, they obey the Lie algebra%
\end{subequations}
\begin{align}
\mathrm{i}\left[  \mathbf{\Sigma}_{X|ab},\mathbf{\Sigma}_{X|cd}\right]   &
=\eta_{ac}\mathbf{\Sigma}_{X|bd}-\eta_{ad}\mathbf{\Sigma}_{X|bc}\nonumber\\
&  -\eta_{bc}\mathbf{\Sigma}_{X|ad}+\eta_{bd}\mathbf{\Sigma}_{X|ac},
\label{Eq:GensLieAlgebra}%
\end{align}
so that $\mathbf{\Sigma}_{L|ab}$ and $\mathbf{\Sigma}_{R|ab}$ constitute
generators of two different, but closely related (by complex conjugation),
four-dimensional spin $\frac{1}{2}$ representations of $\mathrm{Spin}\left(
3,1\right)  $, because $\frac{1}{2}\mathbf{\Sigma}^{L|ij}\mathbf{\Sigma
}_{L|ij}=\frac{1}{2}\mathbf{\Sigma}^{R|ij}\mathbf{\Sigma}_{R|ij}=\frac{3}%
{4}\mathbf{1}_{4}$.

\subsection{Generators of $\mathrm{Spin}\left(  3,1\right)  \times
\mathrm{SU}\left(  2\right)  \times\mathrm{U}\left(  1\right)  $%
\label{Sec:GensAll}}

Eq. (\ref{Eq:GamsCommute}) implies that
\begin{subequations}
\label{Eq:GamsGensCommute}%
\begin{align}
\mathbf{0}_{4} &  =\left[  \mathbf{\Gamma}_{L|a},\mathbf{\Sigma}%
_{R|cd}\right]  ,\label{Eq:GamsGensCommuteLR}\\
\mathbf{0}_{4} &  =\left[  \mathbf{\Gamma}_{R|a},\mathbf{\Sigma}%
_{L|cd}\right]  .\label{Eq:GamsGensCommuteRL}%
\end{align}
In conjunction with Eq. (\ref{Eq:GamsCommute}), these relations imply that
(note the crossing of the $L$ and $R$ sectors) $\mathbf{\Sigma}_{L|cd}$ and
$\mathbf{\Gamma}_{R|a}$ together, and $\mathbf{\Sigma}_{R|cd}$ and
$\mathbf{\Gamma}_{L|a}$ together constitute two different, but closely related
(by complex conjugation), four-dimensional representations of $\mathrm{Spin}%
\left(  3,1\right)  \times\mathrm{SU}\left(  2\right)  \times\mathrm{U}\left(
1\right)  $.

\subsection{Generators of $\mathrm{SO}\left(  3,1\right)  $}

Define the matrices $\mathbf{\Sigma}_{V|ab}\in\mathrm{M}\left(  4,\mathbb{R}%
\right)  $ by%
\end{subequations}
\begin{equation}
\mathrm{i}\left(  \mathbf{\Sigma}_{V|ab}\right)  _{cd}=\eta_{ac}\eta_{bd}%
-\eta_{ad}\eta_{bc}. \label{Eq:GensDefV}%
\end{equation}
They obey the exact same Lie algebra as do $\mathbf{\Sigma}_{L|ab}$ and
$\mathbf{\Sigma}_{R|ab}$, i.e., the Lie algebra given by Eq.
(\ref{Eq:GensLieAlgebra}) with $\mathbf{\Sigma}_{X|ab}$ replaced by
$\mathbf{\Sigma}_{V|ab}$, so they constitute generators of the vector
representation of $\mathrm{SO}\left(  3,1\right)  $, because $\frac{1}%
{2}\left(  \mathbf{\Sigma}^{V|ij}\mathbf{\Sigma}_{V|ij}\right)  _{kl}=2\left(
\mathbf{1}_{3}\right)  _{kl}$.

The matrices $\mathbf{\Gamma}_{X|a}$ and $\mathbf{\Sigma}_{X|ab}$ are related
to $\mathbf{\Sigma}_{V|ab}$ by the relations
\begin{subequations}
\label{Eq:DoubleCover}%
\begin{align}
-\left(  \mathbf{\Sigma}_{V|ab}\right)  ^{c}{}_{d}\mathbf{\Gamma}^{X|d}  &
=\mathbf{\Sigma}_{X|ab}^{\dagger}\mathbf{\Gamma}^{X|c}-\mathbf{\Gamma}%
^{X|c}\mathbf{\Sigma}_{X|ab},\label{Eq:DoubleCoverRaised}\\
+\left(  \mathbf{\Sigma}_{V|ab}\right)  ^{d}{}_{c}\mathbf{\Gamma}_{X|d}  &
=\mathbf{\Sigma}_{X|ab}^{\dagger}\mathbf{\Gamma}_{X|c}-\mathbf{\Gamma}%
_{X|c}\mathbf{\Sigma}_{X|ab}. \label{Eq:DoubleCoverLowered}%
\end{align}

\subsection{Transformations and invariants}

Define the matrices $\mathbf{\Lambda}_{L},\mathbf{\Lambda}_{R}\in
\mathrm{M}\left(  4,\mathbb{C}\right)  $ and $\mathbf{\Lambda}_{V}%
\in\mathrm{M}\left(  4,\mathbb{R}\right)  $ by%
\end{subequations}
\begin{align*}
\mathbf{\Lambda}_{L}  &  =\exp\left(  -\frac{\mathrm{i}}{2}\theta
^{ab}\mathbf{\Sigma}_{L|ab}\right)  ,\\
\mathbf{\Lambda}_{R}  &  =\exp\left(  -\frac{\mathrm{i}}{2}\theta
^{ab}\mathbf{\Sigma}_{R|ab}\right)  ,\\
\mathbf{\Lambda}_{V}  &  =\exp\left(  -\frac{\mathrm{i}}{2}\theta
^{ab}\mathbf{\Sigma}_{V|ab}\right)  ,
\end{align*}
where $\theta_{ab}=-\theta_{ba}\in\mathbb{R}$. From Eq.
(\ref{Eq:GensInvolution}) it straightforwardly follows that%
\begin{align*}
\mathbf{\Lambda}_{L}^{\ast}  &  =\mathbf{\Lambda}_{R},\quad\mathbf{\Lambda
}_{R}^{\ast}=\mathbf{\Lambda}_{L},\\
\mathbf{\Lambda}_{L}^{\mathrm{T}}  &  =\mathbf{\eta\Lambda}_{L}^{-1}%
\mathbf{\eta,\quad\Lambda}_{R}^{\mathrm{T}}=\mathbf{\eta\Lambda}_{R}%
^{-1}\mathbf{\eta,}\\
\mathbf{\Lambda}_{L}^{\dagger}  &  =\mathbf{\eta\Lambda}_{R}^{-1}%
\mathbf{\eta,\quad\Lambda}_{R}^{\dagger}=\mathbf{\eta\Lambda}_{L}%
^{-1}\mathbf{\eta.}%
\end{align*}

\begin{remark}
\normalfont Note that under transposition, $\mathbf{\Lambda}_{L}$ and
$\mathbf{\Lambda}_{R}$ surprisingly behave exactly as does $\mathbf{\Lambda
}_{V}$, which obeys $\mathbf{\Lambda}_{L}^{\mathrm{T}}=\mathbf{\eta\Lambda
}_{V}^{-1}\mathbf{\eta}$, the relation responsible for the invariance of the
line element in the special theory of relativity.
\end{remark}

From Eqs. (\ref{Eq:DoubleCover}) it follows that%
\begin{align*}
\left(  \mathbf{\Lambda}_{V}\right)  ^{a}{}_{b}\mathbf{\Gamma}^{X|b}  &
=\mathbf{\Lambda}_{X}^{\dagger}\mathbf{\Gamma}^{X|a}\mathbf{\Lambda}_{X},\\
\left(  \mathbf{\Lambda}_{V}^{-1}\right)  ^{b}{}_{a}\mathbf{\Gamma}_{X|b}  &
=\mathbf{\Lambda}_{X}^{\dagger}\mathbf{\Gamma}_{X|a}\mathbf{\Lambda}_{X}.
\end{align*}
These relations imply, that if $\mathbf{\psi}_{X}$ are two four-spinors
transforming as $\mathbf{\psi}_{X}^{\prime}=\mathbf{\Lambda}_{X}\mathbf{\psi
}_{X}$ under a Lorentz transformation, then $\mathbf{\psi}_{X}^{\dagger
}\mathbf{\Gamma}^{X|a}\partial_{a}\mathbf{\psi}_{X}$ are invariants, because
$\partial_{a}$ transforms as $\partial_{a}^{\prime}=\left(  \mathbf{\Lambda
}_{V}^{-1}\right)  ^{b}{}_{a}\partial_{b}$, and $\mathbf{\psi}_{L}^{\dagger
}\mathbf{\eta\psi}_{R}$ and $\mathbf{\psi}_{R}^{\dagger}\mathbf{\eta\psi}_{L}$
(note the surprising appearance of $\mathbf{\eta}$) are invariants.

\section{Lagrangian\label{Section:Lagrangian}}

Consider the Lagrangian (note the explicit appearance of $\mathbf{\eta}$ in
the mass terms)%
\begin{equation}
\mathcal{L}=\mathbf{\Psi}^{\dagger}\left(
\begin{array}
[c]{cc}%
e^{\mu}{}_{a}\mathbf{\Gamma}^{L|a}\mathbf{D}_{L|\mu} & m^{\ast}\mathbf{\eta}\\
m\mathbf{\eta} & e^{\mu}{}_{a}\mathbf{\Gamma}^{R|a}\mathbf{D}_{R|\mu}%
\end{array}
\right)  \mathbf{\Psi}+\mathrm{h.c.}, \label{Eq:Lagrangian}%
\end{equation}
where (note for the inner interactions $\mathbf{G}_{X|\mu}^{\mathrm{inner}}$
the crossing of the $L$ and $R$ sectors)%
\begin{align*}
\mathbf{D}_{X|\mu}  &  =\mathbf{1}_{4}\partial_{\mu}+\mathbf{G}_{X|\mu
}^{\mathrm{outer}}+\mathbf{G}_{X|\mu}^{\mathrm{inner}},\\
\mathbf{G}_{X|\mu}^{\mathrm{outer}}  &  =\frac{1}{2}\omega_{\mu}{}%
^{ab}\mathbf{\Sigma}_{X|ab},\\
\mathbf{G}_{L|\mu}^{\mathrm{inner}}  &  =\frac{\mathrm{i}}{\hbar}gt_{L}W^{i}%
{}_{\mu}\left(  \mathrm{i}\mathbf{\Gamma}_{R|i}\right)  +\frac{\mathrm{i}%
}{\hbar}\frac{g^{\prime}}{2}y_{L}B_{\mu}\left(  \mathrm{i}\mathbf{\Gamma
}_{R|0}\right)  ,\\
\mathbf{G}_{R|\mu}^{\mathrm{inner}}  &  =\frac{\mathrm{i}}{\hbar}gt_{R}W^{i}%
{}_{\mu}\left(  \mathrm{i}\mathbf{\Gamma}_{L|i}\right)  +\frac{\mathrm{i}%
}{\hbar}\frac{g^{\prime}}{2}y_{R}B_{\mu}\left(  \mathrm{i}\mathbf{\Gamma
}_{L|0}\right)  .
\end{align*}
The fields are: An eight-spinor field $\mathbf{\Psi}^{\mathrm{T}}\equiv\left(
\mathbf{\psi}_{L}^{\mathrm{T}},\mathbf{\psi}_{R}^{\mathrm{T}}\right)  $, a
vierbein field $e^{a}{}_{\mu}$ and its associated minimal spin connection
$\omega_{\mu}{}^{ab}=g^{\rho\sigma}e^{a}{}_{\rho}\nabla_{\mu}e^{b}{}_{\sigma}%
$, see Ref. \cite[Sec. 31.A]{Weinberg}, and $\mathrm{SU}\left(  2\right)  $
and $\mathrm{U}\left(  1\right)  $ gauge fields $W^{i}{}_{\mu}$ and $B_{\mu}$,
respectively. The constants are: Two real masses $m_{1},m_{2}\in\mathbb{R}$
combined into a complex mass $m\equiv m_{1}+\mathrm{i}m_{2}$, coupling
constants $g$ and $g^{\prime}$ for $\mathrm{SU}\left(  2\right)  $ and
$\mathrm{U}\left(  1\right)  $, respectively, and charges $t_{X}$ and $y_{L}$
for $\mathrm{SU}\left(  2\right)  $ and $\mathrm{U}\left(  1\right)  $,
respectively, where $t_{X}=0$ corresponds to an $\mathrm{SU}\left(  2\right)
$ singlet, and $t_{X}=\frac{1}{2}$ corresponds to an $\mathrm{SU}\left(
2\right)  $ doublet. The factor $\frac{1}{2}$ in connection with $g^{\prime}$
is present to be consistent with conventions \cite[p. 428]{Aitchison and Hey}.

Due to the results of Sec. \ref{Section:Setup}, the Lagrangian is
$\mathrm{Spin}\left(  3,1\right)  \times\mathrm{SU}\left(  2\right)
\times\mathrm{U}\left(  1\right)  $ gauge invariant, and it describes an
eight-spinor field $\mathbf{\Psi}$ coupled to the external fields $\omega
_{\mu}{}^{ab}$, and $W^{i}{}_{\mu}$ and $B_{\mu}$.

\begin{remark}
\normalfont In Eq. (\ref{Eq:Lagrangian}), hermitian conjugation $\mathrm{h.c.}%
$ effectively applies to only the terms of the Lagrangian arising from
$\mathbf{1}_{4}\partial_{\mu}$ and $\mathbf{G}_{X|\mu}^{\mathrm{outer}}$,
because the terms arising from $\mathbf{G}_{X|\mu}^{\mathrm{inner}}$ are
hermitian due to Eqs. (\ref{Eq:GamsHermitianConj}) and (\ref{Eq:GamsCommute}),
and the mass terms that couple $\mathbf{\psi}_{L}$ and $\mathbf{\psi}_{R}$ are
each others hermitian conjugate.
\end{remark}

\section{Discussion\label{Section:Discussion}}

A notable feature of the Lagrangian, Eq. (\ref{Eq:Lagrangian}), is that the
$\mathbf{\Gamma}_{X|a}$'s appearing in front of the (covariant) derivatives,
as do the usual Dirac gamma matrices, also appear, although crossed in the $L$
and $R$ sense, in $\mathbf{G}_{X|\mu}^{\mathrm{inner}}$ as $\mathrm{SU}\left(
2\right)  \times\mathrm{U}\left(  1\right)  $ generators. Is that profound?

Define the matrices $\mathbf{P}_{X|e},\mathbf{P}_{X|\nu}\in\mathrm{M}\left(
4,\mathbb{C}\right)  $ by (note the sign difference between the $L$ and $R$
sectors)%
\begin{align*}
\mathbf{P}_{L|e} &  =-\frac{\mathrm{i}}{2}\left(  \mathbf{\Gamma}%
_{R|0}-\mathbf{\Gamma}_{R|3}\right)  ,\\
\mathbf{P}_{L|\nu} &  =-\frac{\mathrm{i}}{2}\left(  \mathbf{\Gamma}%
_{R|0}+\mathbf{\Gamma}_{R|3}\right)  ,\\
\mathbf{P}_{R|e} &  =-\frac{\mathrm{i}}{2}\left(  \mathbf{\Gamma}%
_{L|0}+\mathbf{\Gamma}_{L|3}\right)  ,\\
\mathbf{P}_{R|\nu} &  =-\frac{\mathrm{i}}{2}\left(  \mathbf{\Gamma}%
_{L|0}-\mathbf{\Gamma}_{L|3}\right)  .
\end{align*}
They obey $\mathbf{P}_{X|e}^{2}=\mathbf{P}_{X|e}$ and $\mathbf{P}_{X|\nu}%
^{2}=\mathbf{P}_{X|\nu}$, and $\mathbf{1}_{4}=\mathbf{P}_{X|e}+\mathbf{P}%
_{X|\nu}$ and $\mathbf{0}_{4}=\mathbf{P}_{X|e}\mathbf{P}_{X|\nu}%
=\mathbf{P}_{X|\nu}\mathbf{P}_{X|e}$, so they are projection operators in the
$L$ and $R$ sector, respectively. Because of Eqs. (\ref{Eq:GamsLieAlgebraX})
and (\ref{Eq:GamsCommute}), $\mathbf{P}_{L|e}$ and $\mathbf{P}_{L|\nu}$
commute with all terms in $\mathbf{D}_{L|\mu}$ except the terms arising from
$\mathbf{\Sigma}_{L|ab}$, and $\mathbf{\Gamma}_{R|1}$ and $\mathbf{\Gamma
}_{R|2}$. Analogously for $\mathbf{P}_{R|e}$ and $\mathbf{P}_{R|\nu}$. So, in
the light of Eqs. (\ref{Eq:GamsLieAlgebraL})-(\ref{Eq:GamsLieAlgebraR}) the
matrices $\mathbf{P}_{X|e}$ and $\mathbf{P}_{X|\nu}$ may be considered weak
isospin projection operators, a fact from which their subscripts $e$ and $\nu
$, referring to the electron and neutrino, respectively, are derived from.
Furthermore, because of the sign difference between the $L$ and $R$ sectors,
most importantly (otherwise the mass terms would couple the different isospin
components) they obey%
\begin{align*}
\mathbf{0}_{4} &  =\mathbf{P}_{L|e}\mathbf{\eta P}_{R|\nu}=\mathbf{P}_{L|\nu
}\mathbf{\eta P}_{R|e},\\
\mathbf{0}_{4} &  =\mathbf{P}_{R|e}\mathbf{\eta P}_{L|\nu}=\mathbf{P}_{R|\nu
}\mathbf{\eta P}_{L|e}.
\end{align*}
Therefore, defining the four four-spinors $\mathbf{\psi}_{X|e}=\mathbf{P}%
_{X|e}\mathbf{\psi}_{X}$ and $\mathbf{\psi}_{X|\nu}=\mathbf{P}_{X|\nu
}\mathbf{\psi}_{X}$, the mass terms of the Lagrangian may be written as%
\[
\operatorname{Re}\left(  m^{\ast}\mathbf{\psi}_{L|e}^{\dagger}\mathbf{\eta
\psi}_{R|e}\right)  +\operatorname{Re}\left(  m^{\ast}\mathbf{\psi}_{L|\nu
}^{\dagger}\mathbf{\eta\psi}_{R|\nu}\right)  .
\]
What significance, if any, is there to the explicit appearance of
$\mathbf{\eta}$, comparing it with the usual $2D$-block diagonal
$\mathbf{\gamma}^{0}$ in the mass term of the Dirac Lagrangian? Could the
non-$2D$-block diagonal form of $\mathbf{\eta}$, singling out one of four
components, be connected with the missing component of the neutrino? And
generally, is it any improvement that the usual Dirac projection operators
$\frac{1}{2}\left(  \mathbf{1}_{4}\pm\mathbf{\gamma}_{5}\right)  $ are not present?

On a more speculative note, what happens when the complexified quaternions,
here considered, is (almost irresistibly) generalized to the complexified
octonions? Mathematically, there are some very compelling reasons for such a generalization:

\begin{enumerate}
\item The set of complexified quaternions is a natural subset of the set of
complexified octonions, as the former can be embedded into the latter in
numerous ways.

\item The proofs of Appendix \ref{Appendix:Proofs}, with the sole exception
being the proof of Eq. (\ref{Eq:GamsCommute}), which relies on associativity,
a property the octonions does not have, carry over without any change for
matrices $\mathbf{\Gamma}_{X|A}$ and $\mathbf{\Sigma}_{X|AB}$ (replacing
$\mathbf{\Gamma}_{X|a}$ and $\mathbf{\Sigma}_{X|ab}$ considered in this
article) defined by
\begin{align*}
\left(  \mathbf{\Gamma}_{L|A}\right)  _{CD} &  =\left\langle \mathrm{e}%
_{C},\mathrm{e}_{A}\mathrm{e}_{D}\right\rangle ,\\
\left(  \mathbf{\Gamma}_{R|A}\right)  _{CD} &  =\left\langle \mathrm{e}%
_{C},\mathrm{e}_{D}\mathrm{e}_{A}\right\rangle ,
\end{align*}
and [generators of the spinor representations of $\mathrm{Spin}\left(
7,1\right)  $]%
\begin{align*}
4\mathrm{i}\left(  \mathbf{\Sigma}_{L|AB}\right)  _{CD} &  =\left\langle
\mathrm{e}_{A}\mathrm{e}_{C},\mathrm{e}_{B}\mathrm{e}_{D}\right\rangle
-\left\langle \mathrm{e}_{A}\mathrm{e}_{D},\mathrm{e}_{B}\mathrm{e}%
_{C}\right\rangle ,\\
4\mathrm{i}\left(  \mathbf{\Sigma}_{R|AB}\right)  _{CD} &  =\left\langle
\mathrm{e}_{C}\mathrm{e}_{A},\mathrm{e}_{D}\mathrm{e}_{B}\right\rangle
-\left\langle \mathrm{e}_{D}\mathrm{e}_{A},\mathrm{e}_{C}\mathrm{e}%
_{B}\right\rangle ,
\end{align*}
where $\mathrm{e}_{A}=\left(  \mathrm{i},\mathrm{e}_{I}\right)  \in
\mathbb{C}\otimes\mathbb{O}$ is a basis for the complexified octonions:
$\mathrm{e}_{I}$ are the seven imaginary units of $\mathbb{O}$, obeying
$\mathrm{e}_{I}\mathrm{e}_{J}=-\delta_{IJ}+\psi_{IJ}{}^{K}\mathrm{e}_{K}$,
where $\psi_{IJK}$ are the octonionic structure constants, see for instance
Refs. \cite{Gunaydin and Gursey,Dundarer and Gursey,Dundarer Gursey and
Tze,Bakas et al.}. Of course, various other replacements must be made, for
instance replacing $\mathbf{\eta}\in\mathrm{M}\left(  4,\mathbb{R}\right)  $
by the eight-dimensional Minkowski metric $\mathbf{\eta}_{8}\in\mathrm{M}%
\left(  8,\mathbb{R}\right)  $. Might the requirement of associativity in the
proof of Eq. (\ref{Eq:GamsCommute}), which holds for the complexified
quaternions, but not for the complexified octonions, be the explanation for
the four-dimensionality of spacetime, somehow forcing a $\mathbb{C}%
\otimes\mathbb{H}$-fibration of $\mathbb{C}\otimes\mathbb{O}$?

\item The quaternions and octonions share a \textit{unique} property, although
not utilized in this article: They allow the definition of triple cross
products $X_{L},X_{R}:\left(  \mathbb{C}\otimes\mathbb{D}\right)
^{3}\rightarrow\mathbb{C}\otimes\mathbb{D}$ (where $\mathbb{D}$ denotes either
$\mathbb{H}$ or $\mathbb{O}$) by%
\begin{align*}
3!X_{L}\left(  x,y,z\right)   &  =x\left(  \overline{y}z-\overline{z}y\right)
+\text{cyclic perm},\\
3!X_{R}\left(  x,y,z\right)   &  =\left(  x\overline{y}-y\overline{x}\right)
z+\text{cyclic perm}.
\end{align*}
The cross products $X_{L}$ and $X_{R}$ possess both the orthogonality property
and the (generalized) Pythagorean property \cite{Lounesto},%
\begin{align*}
0  &  =\left\langle X\left(  x_{1},x_{2},x_{3}\right)  ,x_{i}\right\rangle ,\\
\det\left(  \left\langle x_{i},x_{j}\right\rangle \right)   &  =\left\langle
X\left(  x_{1},x_{2},x_{3}\right)  ,X\left(  x_{1},x_{2},x_{3}\right)
\right\rangle ,
\end{align*}
where the suppressed subscript means that the relations apply to both $L$ and
$R$. Trilinear cross products possessing both these properties exist
\textit{only} over algebras of real (or complex) dimension $4$ or $8$, see
Refs. \cite{Lounesto,Zvengrowski}, the underlying reason being the existence
of precisely the division algebras $\mathbb{H}$ and $\mathbb{O}$.

\item The seemingly insignificant relation%
\[
\varepsilon_{abcd}=\mathrm{i}\left\langle X\left(  \mathrm{e}_{a}%
,\mathrm{e}_{b},\mathrm{e}_{c}\right)  ,\mathrm{e}_{d}\right\rangle ,
\]
links duality in four-dimensional spacetime, as controlled by $\varepsilon
_{abcd}$, with two natural structures of the (complex) quaternions, the inner
product and the cross product, as defined above. This relation may be
straightforwardly generalized to%
\begin{align*}
\chi_{L|ABCD} &  =\mathrm{i}\left\langle X_{L}\left(  \mathrm{e}%
_{A},\mathrm{e}_{B},\mathrm{e}_{C}\right)  ,\mathrm{e}_{D}\right\rangle ,\\
\chi_{R|ABCD} &  =\mathrm{i}\left\langle X_{R}\left(  \mathrm{e}%
_{A},\mathrm{e}_{B},\mathrm{e}_{C}\right)  ,\mathrm{e}_{D}\right\rangle ,
\end{align*}
where $\chi_{L|ABCD}$ and $\chi_{R|ABCD}$ are nonequal because of the
nonassociativity of the (complexified) octonions. These structure constants
$\chi_{L|ABCD}$ and $\chi_{R|ABCD}$ allow for the definition of self-duality
in eight-dimensional spacetime of rank \textit{two} tensors:%
\[
T_{AB}=\frac{\mathrm{i}}{2}\lambda_{X}\chi_{X|ABCD}T^{CD}.
\]
It can be shown that the eigenvalues are $\lambda_{L}\in\left\{
+1,-1/3\right\}  $ and $\lambda_{R}\in\left\{  -1,+1/3\right\}  $. In the
quaternionic case the eigenvalues are $\pm1$, as is well-known. Is the
appearance of $\pm1/3$ in the octonionic case somehow related to fractional
(hyper)charges of the quarks?
\end{enumerate}

It is the hope that some or all of these issues will be resolved in the near future.

\appendix

\section{\label{Appendix:Identities}Identities}

The following Lemma lists some useful identities for composition algebras, a
class to which the complexified quaternions belong, see for instance
\cite{Okubo} or \cite{Springer and Veldkamp}. Note, though, that the
normalization of the inner product in \cite{Okubo} and \cite{Springer and
Veldkamp} differ by a factor of $2$. The normalization used in Eq.
(\ref{Eq:ipDef}) is the normalization used in \cite{Okubo}. However,
\cite{Springer and Veldkamp} is mentioned because its overall presentation is
clearer than that of \cite{Okubo}, and as such may be valuable to the reader.

\begin{lemma}
[See \cite{Okubo} or \cite{Springer and Veldkamp}]The following identities
hold for any composition algebra:%
\begin{align}
\left\langle x,y\right\rangle  &  \equiv\left\langle y,x\right\rangle
,\label{Eq:ipSym}\\
\left\langle x,y\right\rangle  &  \equiv\left\langle \overline{x},\overline
{y}\right\rangle , \label{Eq:ipConj}%
\end{align}
and%
\begin{align}
\left\langle x,yz\right\rangle  &  \equiv\left\langle \overline{y}%
x,z\right\rangle ,\label{Eq:ipMoveL}\\
\left\langle xy,z\right\rangle  &  \equiv\left\langle x,z\overline
{y}\right\rangle , \label{Eq:ipMoveR}%
\end{align}
and%
\begin{align}
x\left(  \overline{y}z\right)  +\left(  y\overline{x}\right)  z  &
\equiv2\left\langle x,y\right\rangle z,\label{Eq:ipSumL}\\
\left(  x\overline{y}\right)  z+\left(  x\overline{z}\right)  y  &
\equiv2\left\langle y,z\right\rangle x. \label{Eq:ipSumR}%
\end{align}

\end{lemma}

\section{\label{Appendix:Proofs}Proofs}

Throughout this section the identities of the Lemma of Appendix
\ref{Appendix:Identities} will be used without being explicitly referred to.
Although the equations being proved below could reasonably simply be checked
by explicit numerical calculation, by first calculating explicitly the
four-dimensional matrices $\mathbf{\Gamma}_{L|a}$ and $\mathbf{\Gamma}_{R|a}$,
and $\mathbf{\Sigma}_{L|ab}$ and $\mathbf{\Sigma}_{R|ab}$, using Eqs.
(\ref{Eq:GamsDef}) and (\ref{Eq:GensDef}), the main purpose of presenting
analytical proofs is that the majority of these as stated, the sole exception
being the proof of Eq. (\ref{Eq:GamsCommute}), apply to any composition
algebra, and therefore in particular to both the complexified quaternions and
the complexified octonions.

\subsection{Proof of Eq. (\ref{Eq:GamsInvolution})}

By direct calculation, using $\mathrm{e}_{a}^{\ast}=-\overline{\mathrm{e}}%
_{a}$ and $\overline{\mathrm{e}}_{a}=-\delta_{ab}\mathrm{e}^{b}$,
respectively:%
\begin{align*}
\left(  \mathbf{\Sigma}_{L|a}^{\ast}\right)  _{cd}  &  =\left[  \left(
\mathbf{\Sigma}_{L|a}\right)  _{cd}\right]  ^{\ast}=\left\langle
\mathrm{e}_{c},\mathrm{e}_{a}\mathrm{e}_{d}\right\rangle ^{\ast}=\left\langle
\mathrm{e}_{c}^{\ast},\mathrm{e}_{a}^{\ast}\mathrm{e}_{d}^{\ast}\right\rangle
\\
&  =-\left\langle \overline{\mathrm{e}}_{c},\overline{\mathrm{e}}_{a}%
\overline{\mathrm{e}}_{d}\right\rangle =-\left\langle \mathrm{e}%
_{c},\mathrm{e}_{d}\mathrm{e}_{a}\right\rangle =-\left(  \mathbf{\Sigma}%
_{R|a}\right)  _{cd},
\end{align*}
and%
\begin{align*}
\left(  \mathbf{\Sigma}_{L|a}^{\mathrm{T}}\right)  ^{c}{}_{d}  &  =\delta
^{ce}\left(  \mathbf{\Sigma}_{L|a}\right)  ^{f}{}_{e}\delta_{fd}=\delta
^{ce}\left\langle \mathrm{e}^{f},\mathrm{e}_{a}\mathrm{e}_{e}\right\rangle
\delta_{fd}\\
&  =\left\langle \left(  -\delta_{df}\mathrm{e}^{f}\right)  ,\mathrm{e}%
_{a}\left(  -\delta^{ce}\mathrm{e}_{e}\right)  \right\rangle =\left\langle
\overline{\mathrm{e}}_{d},\mathrm{e}_{a}\overline{\mathrm{e}}^{c}\right\rangle
\\
&  =\left\langle \overline{\mathrm{e}}_{d}\mathrm{e}^{c},\mathrm{e}%
_{a}\right\rangle =\left\langle \mathrm{e}^{c},\mathrm{e}_{d}\mathrm{e}%
_{a}\right\rangle =\left(  \mathbf{\Sigma}_{R|a}\right)  ^{c}{}_{d}.
\end{align*}
The remaining assertion, Eq. (\ref{Eq:GamsHermitianConj}), readily follows
from the matrix identity $\mathbf{M}^{\dagger}\equiv\left(  \mathbf{M}^{\ast
}\right)  ^{\mathrm{T}}\equiv\left(  \mathbf{M}^{\mathrm{T}}\right)  ^{\ast}$.

\subsection{Proof of Eq. (\ref{Eq:GamsCommute})}

Using the completeness relation $\left\langle x,\mathrm{e}_{a}\right\rangle
\left\langle \mathrm{e}^{a},y\right\rangle \equiv\left\langle x,y\right\rangle
$:%
\begin{align*}
\left[  \mathbf{\Sigma}_{L|a}\right]  ^{c}{}_{e}\left[  \mathbf{\Sigma}%
_{R|b}\right]  ^{e}{}_{d}  &  =\left\langle \mathrm{e}^{c},\mathrm{e}%
_{a}\mathrm{e}_{e}\right\rangle \left\langle \mathrm{e}^{e},\mathrm{e}%
_{d}\mathrm{e}_{b}\right\rangle =\left\langle \overline{\mathrm{e}}%
_{a}\mathrm{e}^{c},\mathrm{e}_{d}\mathrm{e}_{b}\right\rangle ,\\
\left[  \mathbf{\Sigma}_{R|b}\right]  ^{c}{}_{e}\left[  \mathbf{\Sigma}%
_{L|a}\right]  ^{e}{}_{d}  &  =\left\langle \mathrm{e}^{c},\mathrm{e}%
_{e}\mathrm{e}_{b}\right\rangle \left\langle \mathrm{e}^{e},\mathrm{e}%
_{a}\mathrm{e}_{d}\right\rangle =\left\langle \mathrm{e}^{c}\overline
{\mathrm{e}}_{b},\mathrm{e}_{a}\mathrm{e}_{d}\right\rangle .
\end{align*}
These two expressions are equal because the (complexified) quaternions are
\textit{associative} (a property which breaks down when generalizing to
complexified octonions) so that%
\begin{align*}
\left\langle \overline{\mathrm{e}}_{a}\mathrm{e}^{c},\mathrm{e}_{d}%
\mathrm{e}_{b}\right\rangle  &  =\left\langle \mathrm{e}^{c},\mathrm{e}%
_{a}\left(  \mathrm{e}_{d}\mathrm{e}_{b}\right)  \right\rangle \\
&  =\left\langle \mathrm{e}^{c},\left(  \mathrm{e}_{a}\mathrm{e}_{d}\right)
\mathrm{e}_{b}\right\rangle \\
&  =\left\langle \mathrm{e}^{c}\overline{\mathrm{e}}_{b},\mathrm{e}%
_{a}\mathrm{e}_{d}\right\rangle .
\end{align*}

\subsection{Proof of Eqs. (\ref{Eq:GamsEtaAntiCom}) and (\ref{Eq:GensEtaCom})}

Only the proof for $L$ will be given, as the proof for $R$ is completely
analogous. Eq. (\ref{Eq:GamsEtaAntiCom}) is proved as follows:%
\begin{align*}
\left(  \mathbf{\Sigma}_{L|a}^{\eta}\right)  ^{c}{}_{d}  &  =\left(
\mathbf{\eta\Sigma}_{L|a}^{\mathrm{T}}\mathbf{\eta}\right)  ^{c}{}_{d}=\left(
\mathbf{\eta}\right)  ^{c}{}_{e}\left(  \mathbf{\Sigma}_{L|a}^{\mathrm{T}%
}\right)  ^{e}{}_{f}\left(  \mathbf{\eta}\right)  ^{f}{}_{d}\\
&  =\eta^{ce}\left(  \mathbf{\Sigma}_{L|a}\right)  ^{f}{}_{e}\eta_{fd}%
=\eta^{ce}\left\langle \mathrm{e}^{f},\mathrm{e}_{a}\mathrm{e}_{e}%
\right\rangle \eta_{fd}\\
&  =\left\langle \mathrm{e}_{d},\mathrm{e}_{a}\mathrm{e}^{c}\right\rangle ,
\end{align*}
which, using the completeness relation $\left\langle x,\mathrm{e}%
_{a}\right\rangle \left\langle \mathrm{e}^{a},y\right\rangle \equiv
\left\langle x,y\right\rangle $, implies that%
\begin{align*}
\left(  \mathbf{\Sigma}_{L|a}^{\eta}\mathbf{\Sigma}_{L|b}\right)  ^{c}{}_{d}
&  \equiv\left(  \mathbf{\Sigma}_{L|a}^{\eta}\right)  ^{c}{}_{e}\left(
\mathbf{\Sigma}_{L|b}\right)  ^{e}{}_{d}\\
&  =\left\langle \mathrm{e}_{e},\mathrm{e}_{a}\mathrm{e}^{c}\right\rangle
\left\langle \mathrm{e}^{e},\mathrm{e}_{b}\mathrm{e}_{d}\right\rangle \\
&  =\left\langle \mathrm{e}_{a}\mathrm{e}^{c},\mathrm{e}_{b}\mathrm{e}%
_{d}\right\rangle ,
\end{align*}
which implies that%
\begin{align*}
\left(  \left[  \mathbf{\Gamma}_{L|a},\mathbf{\Gamma}_{L|b}\right]  _{\eta
+}\right)  ^{c}{}_{d}  &  =\left(  \mathbf{\Sigma}_{L|a}^{\eta}\mathbf{\Sigma
}_{L|b}+\mathbf{\Sigma}_{L|b}^{\eta}\mathbf{\Sigma}_{L|a}\right)  ^{c}{}_{d}\\
&  =\left\langle \mathrm{e}_{a}\mathrm{e}^{c},\mathrm{e}_{b}\mathrm{e}%
_{d}\right\rangle +\left\langle \mathrm{e}_{b}\mathrm{e}^{c},\mathrm{e}%
_{a}\mathrm{e}_{d}\right\rangle \\
&  =\left\langle \mathrm{e}_{a},\left(  \mathrm{e}_{b}\mathrm{e}_{d}\right)
\overline{\mathrm{e}}^{c}\right\rangle +\left\langle \left(  \mathrm{e}%
_{b}\mathrm{e}^{c}\right)  \overline{\mathrm{e}}_{d},\mathrm{e}_{a}%
\right\rangle \\
&  =\left\langle \mathrm{e}_{a},\left(  \mathrm{e}_{b}\mathrm{e}_{d}\right)
\overline{\mathrm{e}}^{c}+\left(  \mathrm{e}_{b}\mathrm{e}^{c}\right)
\overline{\mathrm{e}}_{d}\right\rangle \\
&  =2\left\langle \mathrm{e}_{a},\mathrm{e}_{b}\right\rangle \left\langle
\overline{\mathrm{e}}^{c},\overline{\mathrm{e}}_{d}\right\rangle =2\eta
_{ab}\left(  \mathbf{1}_{4}\right)  ^{c}{}_{d}.
\end{align*}
Eq. (\ref{Eq:GensEtaCom}) follows directly from the second equation in the
proof above for Eq. (\ref{Eq:GamsEtaAntiCom}), and the defining equation of
$\mathbf{\Sigma}_{L|ab}$ and $\mathbf{\Sigma}_{R|ab}$, Eq. (\ref{Eq:GensDef}).

\subsection{Proof of Eq. (\ref{Eq:GensInvolution})}

Using Eqs. (\ref{Eq:GamsComplexConj}) and (\ref{Eq:GensEtaCom}):%
\begin{align*}
-4\mathrm{i}\mathbf{\Sigma}_{L|ab}^{\ast}  &  =\left(  \mathbf{\Gamma}%
_{L|a}^{\eta}\mathbf{\Gamma}_{L|b}-\mathbf{\Gamma}_{L|b}^{\eta}\mathbf{\Gamma
}_{L|a}\right)  ^{\ast}\\
&  =\mathbf{\eta}\left(  \mathbf{\Gamma}_{L|a}^{\ast}\right)  ^{\mathrm{T}%
}\mathbf{\eta\Gamma}_{L|b}^{\ast}-\mathbf{\eta}\left(  \mathbf{\Gamma}%
_{L|b}^{\ast}\right)  ^{\mathrm{T}}\mathbf{\eta\Gamma}_{L|a}^{\ast}\\
&  =\mathbf{\eta\Gamma}_{R|a}^{\mathrm{T}}\mathbf{\eta\Gamma}_{R|b}%
-\mathbf{\eta\Gamma}_{R|b}^{\mathrm{T}}\mathbf{\eta\Gamma}_{R|a}\\
&  =4\mathrm{i}\mathbf{\Sigma}_{R|ab}.
\end{align*}
Using Eq. (\ref{Eq:GensEtaCom}):%
\begin{align*}
4\mathrm{i}\mathbf{\Sigma}_{X|ab}^{\mathrm{T}}  &  =\left(  \mathbf{\Gamma
}_{X|a}^{\eta}\mathbf{\Gamma}_{X|b}-\mathbf{\Gamma}_{X|b}^{\eta}%
\mathbf{\Gamma}_{X|a}\right)  ^{\mathrm{T}}\\
&  =\mathbf{\Gamma}_{X|b}^{\mathrm{T}}\mathbf{\eta\Gamma}_{X|a}\mathbf{\eta
}-\mathbf{\Gamma}_{X|a}^{\mathrm{T}}\mathbf{\eta\Gamma}_{X|b}\mathbf{\eta}\\
&  =\mathbf{\eta}\left(  \mathbf{\Gamma}_{X|b}^{\eta}\mathbf{\Gamma}%
_{X|a}-\mathbf{\Gamma}_{X|a}^{\eta}\mathbf{\Gamma}_{X|b}\right)  \mathbf{\eta
}\\
&  =-4\mathrm{i}\mathbf{\eta\Sigma}_{X|ab}\mathbf{\eta}.
\end{align*}
The remaining assertion, Eq. (\ref{Eq:GensHermitianConj}), readily follows
from the matrix identity $\mathbf{M}^{\dagger}\equiv\left(  \mathbf{M}^{\ast
}\right)  ^{\mathrm{T}}\equiv\left(  \mathbf{M}^{\mathrm{T}}\right)  ^{\ast}$.

\subsection{Proof of Eq. (\ref{Eq:GensLieAlgebra})}

To compactify the calculations, the subscript $X|$ has been dropped
throughout. Consider the expression $\mathbf{\Gamma}_{a}^{\eta}\mathbf{\Gamma
}_{b}\mathbf{\Gamma}_{c}^{\eta}\mathbf{\Gamma}_{d}$. Using fourfoldly Eq.
(\ref{Eq:GamsEtaAntiCom}) to move $\mathbf{\Gamma}_{a}^{\eta}\mathbf{\Gamma
}_{b}$ through $\mathbf{\Gamma}_{c}^{\eta}\mathbf{\Gamma}_{d}$, it follows
that%
\begin{align*}
\left[  \mathbf{\Gamma}_{a}^{\eta}\mathbf{\Gamma}_{b},\mathbf{\Gamma}%
_{c}^{\eta}\mathbf{\Gamma}_{d}\right]   &  =2\eta_{ac}\mathbf{\Gamma}%
_{d}^{\eta}\mathbf{\Gamma}_{b}-2\eta_{ad}\mathbf{\Gamma}_{c}^{\eta
}\mathbf{\Gamma}_{b}\\
&  +2\eta_{bc}\mathbf{\Gamma}_{a}^{\eta}\mathbf{\Gamma}_{d}-2\eta
_{bd}\mathbf{\Gamma}_{a}^{\eta}\mathbf{\Gamma}_{c}.
\end{align*}
Using fourfoldly this result in the expression%
\begin{align*}
-16\left[  \mathbf{\Sigma}_{ab},\mathbf{\Sigma}_{cd}\right]   &  =\left[
\left[  \mathbf{\Gamma}_{a},\mathbf{\Gamma}_{b}\right]  _{\eta-},\left[
\mathbf{\Gamma}_{c},\mathbf{\Gamma}_{d}\right]  _{\eta-}\right] \\
&  =\left[  \mathbf{\Gamma}_{a}^{\eta}\mathbf{\Gamma}_{b},\mathbf{\Gamma}%
_{c}^{\eta}\mathbf{\Gamma}_{d}\right]  -\left[  \mathbf{\Gamma}_{a}^{\eta
}\mathbf{\Gamma}_{b},\mathbf{\Gamma}_{d}^{\eta}\mathbf{\Gamma}_{c}\right] \\
&  -\left[  \mathbf{\Gamma}_{b}^{\eta}\mathbf{\Gamma}_{a},\mathbf{\Gamma}%
_{c}^{\eta}\mathbf{\Gamma}_{d}\right]  +\left[  \mathbf{\Gamma}_{b}^{\eta
}\mathbf{\Gamma}_{a},\mathbf{\Gamma}_{d}^{\eta}\mathbf{\Gamma}_{c}\right]  ,
\end{align*}
collecting identical terms, and using again Eq. (\ref{Eq:GamsEtaAntiCom}),
yields%
\begin{align*}
-16\left[  \mathbf{\Sigma}_{ab},\mathbf{\Sigma}_{cd}\right]   &  =-4\eta
_{ac}\left[  \mathbf{\Gamma}_{b},\mathbf{\Gamma}_{d}\right]  _{\eta-}%
+4\eta_{ad}\left[  \mathbf{\Gamma}_{b},\mathbf{\Gamma}_{c}\right]  _{\eta-}\\
&  +4\eta_{bc}\left[  \mathbf{\Gamma}_{a},\mathbf{\Gamma}_{d}\right]  _{\eta
-}-4\eta_{bd}\left[  \mathbf{\Gamma}_{a},\mathbf{\Gamma}_{c}\right]  _{\eta-},
\end{align*}
from which the result follows.

\begin{remark}
\normalfont The proof is completely analogous to the proof of the assertion
that $-\frac{\mathrm{i}}{4}\left[  \mathbf{\gamma}_{a},\mathbf{\gamma}%
_{b}\right]  $, where $\mathbf{\gamma}_{a}$ are the Dirac matrices obeying
$2\eta_{ab}\mathbf{1}_{4}=\left\{  \mathbf{\gamma}_{a},\mathbf{\gamma}%
_{b}\right\}  $, are generators of $\mathrm{Spin}\left(  3,1\right)  $. That
is the main reason for introducing above the (anti)commutator-like brackets
$\left[  \cdot,\cdot\right]  _{\eta\pm}$.
\end{remark}

\subsection{Proof of Eq. (\ref{Eq:GamsGensCommute})}

Only the proof of Eq. (\ref{Eq:GamsGensCommuteLR}) will be given, as the proof
of Eq. (\ref{Eq:GamsGensCommuteRL}) is completely analogous. Using
$\overline{\mathrm{e}}_{a}=-\delta_{ab}\mathrm{e}^{b}$:%
\begin{align*}
\left(  \mathbf{\Sigma}_{L|a}^{\eta}\right)  ^{c}{}_{d}  &  =\left(
\mathbf{\eta\Sigma}_{L|a}^{\mathrm{T}}\mathbf{\eta}\right)  ^{c}{}_{d}=\left(
\mathbf{\eta}\right)  ^{c}{}_{e}\left(  \mathbf{\Sigma}_{L|a}^{\mathrm{T}%
}\right)  ^{e}{}_{f}\left(  \mathbf{\eta}\right)  ^{f}{}_{d}\\
&  =\eta^{ce}\left(  \mathbf{\Sigma}_{L|a}\right)  ^{f}{}_{e}\eta_{fd}%
=\eta^{ce}\left\langle \mathrm{e}^{f},\mathrm{e}_{a}\mathrm{e}_{e}%
\right\rangle \eta_{fd}\\
&  =\left\langle \mathrm{e}_{d},\mathrm{e}_{a}\mathrm{e}^{c}\right\rangle
=\left\langle \overline{\mathrm{e}}_{d},\overline{\mathrm{e}}^{c}%
\overline{\mathrm{e}}_{a}\right\rangle =\left\langle \mathrm{e}^{c}%
,\overline{\mathrm{e}}_{a}\mathrm{e}_{d}\right\rangle \\
&  =-\delta_{ab}\left\langle \mathrm{e}^{c},\overline{\mathrm{e}}%
^{b}\mathrm{e}_{d}\right\rangle =-\delta_{ab}\left(  \mathbf{\Sigma}%
^{L|b}\right)  ^{c}{}_{d},
\end{align*}
which, using Eq. (\ref{Eq:GamsCommute}), implies that%
\begin{align*}
4\mathrm{i}\left[  \mathbf{\Gamma}_{L|a},\mathbf{\Sigma}_{R|cd}\right]   &
=\left[  \mathbf{\Gamma}_{L|a},\left[  \mathbf{\Gamma}_{R|c},\mathbf{\Gamma
}_{R|d}\right]  _{\eta-}\right] \\
&  =\mathbf{\Gamma}_{L|a}\mathbf{\Gamma}_{R|c}^{\eta}\mathbf{\Gamma}%
_{R|d}-\mathbf{\Gamma}_{L|a}\mathbf{\Gamma}_{R|d}^{\eta}\mathbf{\Gamma}%
_{R|c}\\
&  -\mathbf{\Gamma}_{R|c}^{\eta}\mathbf{\Gamma}_{L|a}\mathbf{\Gamma}%
_{R|d}+\mathbf{\Gamma}_{R|d}^{\eta}\mathbf{\Gamma}_{L|a}\mathbf{\Gamma}%
_{R|c}\\
&  =\left(  \mathbf{\Gamma}_{L|a}\mathbf{\Gamma}_{R|c}^{\eta}-\mathbf{\Gamma
}_{R|c}^{\eta}\mathbf{\Gamma}_{L|a}\right)  \mathbf{\Gamma}_{R|d}\\
&  -\left(  \mathbf{\Gamma}_{L|a}\mathbf{\Gamma}_{R|d}^{\eta}-\mathbf{\Gamma
}_{R|d}^{\eta}\mathbf{\Gamma}_{L|a}\right)  \mathbf{\Gamma}_{R|c}\\
&  =-\delta_{ce}\left[  \mathbf{\Gamma}_{L|a},\mathbf{\Gamma}^{R|e}\right]
\mathbf{\Gamma}_{R|d}\\
&  +\delta_{de}\left[  \mathbf{\Gamma}_{L|a},\mathbf{\Gamma}^{R|e}\right]
\mathbf{\Gamma}_{R|c}=\mathbf{0}_{4}.
\end{align*}

\subsection{Proof of Eq. (\ref{Eq:DoubleCover})}

Only the proof of Eq. (\ref{Eq:DoubleCoverRaised}) will be given, as the proof
of Eq. (\ref{Eq:DoubleCoverLowered}) is analogous. Using Eq.
(\ref{Eq:GamsHermitianConj}):%
\[
\left(  \mathbf{\Gamma}_{X|a}^{\eta}\mathbf{\Gamma}_{X|b}\right)  ^{\dagger
}=\mathbf{\Gamma}_{X|b}^{\dagger}\left(  \mathbf{\Gamma}_{X|a}^{\dagger
}\right)  ^{\eta}=\mathbf{\Gamma}_{X|b}\mathbf{\Gamma}_{X|a}^{\eta},
\]
which, using Eqs. (\ref{Eq:GamsEtaAntiCom}) and (\ref{Eq:GensEtaCom}), implies
that (where to compactify the calculations, the subscript $X|$ has been
dropped throughout)%
\begin{align*}
4\mathrm{i}\left(  \mathbf{\Sigma}_{ab}^{\dagger}\mathbf{\Gamma}%
^{c}-\mathbf{\Gamma}^{c}\mathbf{\Sigma}_{ab}\right)   &  =\mathbf{\Gamma}%
_{a}\left[  2\delta_{b}^{c}\mathbf{1}_{4}-\left(  \mathbf{\Gamma}^{c}\right)
^{\eta}\mathbf{\Gamma}_{b}\right]  -\mathbf{\Gamma}^{c}\mathbf{\Gamma}%
_{a}^{\eta}\mathbf{\Gamma}_{b}\\
&  -\mathbf{\Gamma}_{b}\left[  2\delta_{a}^{c}\mathbf{1}_{4}-\left(
\mathbf{\Gamma}^{c}\right)  ^{\eta}\mathbf{\Gamma}_{a}\right]  +\mathbf{\Gamma
}^{c}\mathbf{\Gamma}_{b}^{\eta}\mathbf{\Gamma}_{a}\\
&  =2\delta_{b}^{c}\mathbf{\Gamma}_{a}-\left[  \mathbf{\Gamma}_{a}\left(
\mathbf{\Gamma}^{c}\right)  ^{\eta}+\mathbf{\Gamma}^{c}\mathbf{\Gamma}%
_{a}^{\eta}\right]  \mathbf{\Gamma}_{b}\\
&  -2\delta_{a}^{c}\mathbf{\Gamma}_{b}+\left[  \mathbf{\Gamma}_{b}\left(
\mathbf{\Gamma}^{c}\right)  ^{\eta}+\mathbf{\Gamma}^{c}\mathbf{\Gamma}%
_{b}^{\eta}\right]  \mathbf{\Gamma}_{a}\\
&  =2\delta_{b}^{c}\mathbf{\Gamma}_{a}-\left[  \mathbf{\Gamma}^{c}%
,\mathbf{\Gamma}_{a}\right]  _{\eta+}^{\dagger}\mathbf{\Gamma}_{b}\\
&  -2\delta_{a}^{c}\mathbf{\Gamma}_{b}+\left[  \mathbf{\Gamma}^{c}%
,\mathbf{\Gamma}_{b}\right]  _{\eta+}^{\dagger}\mathbf{\Gamma}_{a}\\
&  =4\left(  \delta_{b}^{c}\eta_{ad}-\delta_{a}^{c}\eta_{bd}\right)
\mathbf{\Gamma}^{d}\\
&  =-4\mathrm{i}\left(  \mathbf{\Sigma}_{V|ab}\right)  ^{c}{}_{d}%
\mathbf{\Gamma}^{d}.
\end{align*}

\end{document}